\documentclass [12pt]{article}
\usepackage{amsmath}
\usepackage{amsfonts}
\usepackage{epsfig}
\begin{document}
\def\b{\bar}
\def\d{\partial}
\def\D{\Delta}
\def\cD{{\cal D}}
\def\cK{{\cal K}}
\def\f{\varphi}
\def\g{\gamma}
\def\G{\Gamma}
\def\l{\lambda}
\def\L{\Lambda}
\def\M{{\Cal M}}
\def\m{\mu}
\def\n{\nu}
\def\p{\psi}
\def\q{\b q}
\def\r{\rho}
\def\t{\tau}
\def\x{\phi}
\def\X{\~\xi}
\def\~{\widetilde}
\def\h{\eta}
\def\bZ{\bar Z}
\def\cY{\bar Y}
\def\bY3{\bar Y_{,3}}
\def\Y3{Y_{,3}}
\def\z{\zeta}
\def\Z{{\b\zeta}}
\def\Y{{\bar Y}}
\def\cZ{{\bar Z}}
\def\`{\dot}
\def\be{\begin{equation}}
\def\ee{\end{equation}}
\def\bea{\begin{eqnarray}}
\def\eea{\end{eqnarray}}
\def\half{\frac{1}{2}}
\def\fn{\footnote}
\def\bh{black hole \ }
\def\cL{{\cal L}}
\def\cH{{\cal H}}
\def\cF{{\cal F}}
\def\cP{{\cal P}}
\def\cM{{\cal M}}
\def\ik{ik}
\def\mn{{\mu\nu}}
\def\a{\alpha}

\title{Complex structure of  Kerr-Schild geometry:
 Calabi-Yau twofold from the Kerr theorem}
\author{Alexander Burinskii \\
 Theor.Phys. Lab., NSI, Russian Academy of Sciences, \\
B. Tulskaya 52 Moscow 115191 Russia\footnote{E-mail address:
bur@ibrae.ac.ru}}

\date{Essay written for the Gravity Research Foundation 2012
Awards for Essays on Gravitation. (March 17, 2012) } \maketitle

\maketitle

\begin{abstract}
 We consider Newman's representation of the Kerr geometry as a complex
 retarded-time construction generated by a source propagating along a complex
 world-line. We notice that the complex world-line forms really an open
 complex string, endpoints of which should have independent dynamics by the
 string excitations. The adjoined to complex Kerr string twistorial
structure is determined by the Kerr theorem, and we obtain that
the resulting Kerr's equation describes a quartic in projective
twistor $CP^3 ,$ which is known as Calabi-Yau twofold of
superstring theory. Along with other remarkable
 similarities with superstring theory, the Kerr geometry has principal
 distinctions being the four-dimensional theory
consistent with gravity at the Compton scale, contrary to the
Planck scale of the superstring theory.
\end{abstract}


\newpage

{\bf Introduction.} The Kerr solution, being  obtained as a metric
of a "rotating body" \cite{Kerr} with angular momentum $J=m|a| ,$
has found widespread application as a metric of rotating black
hole. The parameter $a= J/m$ is radius of the Kerr singular ring.
For $|a|<m ,$ the ring is covered by horizon, but for $|a|>m $ it
is naked, and for the censorship principle, it should be covered
by a source. The long term investigations showed
\cite{Keres,Isr,Lop,BurBag,BurSol},
 that the source of Kerr-Newman (KN) solution forms a rotating
 disk, an oblate bubble (membrane) with flat interior.
 The charged KN solution \cite{KerNew} has found application as a
consistent with gravity classical model of spinning particle,
\cite{Isr,Lop,Car,DKS,Bur0,TN}, which has gyromagnetic ratio $g=2
,$ as that of the Dirac electron \cite{Car,DKS} and displays other
relationships with the Dirac electron,
\cite{DirKer,BurAxi,BurQ,Beyond}, twistor theory
\cite{BurA,BurPreQ,BurTwi}, soliton models
\cite{BurBag,BurSol,Dym} and with basic structures of superstring
theory \cite{BurQ,BurTwi,IvBur,BurSen,BurCStr}. In this note we
consider complex structure of Kerr geometry \cite{BurCStr} and one
new evidence of its relationship with twistors and superstring
theory, namely the following from the Kerr theorem presence of
\emph{the Calabi-Yau twofold as a quartic of the projective
twistor space.}

 KN metric is represented in the Kerr-Schild (KS) form \cite{DKS},
 \be g_\mn=\eta _\mn + 2h
e^3_\m e^3_\n \label{KSh} , \ee where $\eta_\mn$ is auxiliary
Minkowski background in Cartesian coordinates ${\rm x}= x^\m
=(t,x,y,z),$ \be h = P^2 \frac {mr-e^2/2} {r^2 + a^2 \cos^2
\theta}, \quad P=(1+Y\Y)/ \sqrt 2 ,  \label{h}\ee and $e^3 (\rm
x)$ is a tangent direction to a \emph{Principal Null Congruence
(PNC)}, which is determined by the form\fn{Here $ \z =
(x+iy)/\sqrt 2 ,\quad  \Z = (x-iy)/\sqrt 2 , u = (z + t)/\sqrt 2
,\quad v = (z - t)/\sqrt 2, $ are the null Cartesian coordinates,
$r, \theta, \phi $ are the Kerr oblate spheroidal coordinates, and
$Y (\rm x) =e^{i\phi} \tan \frac{\theta}{2} $ is a projective
angular coordinate.  The used signature is $(-+++) $.} \be e^3_\m
dx^\m =du + \bar Y d \zeta + Y d \bar\zeta - Y\bar Y dv ,
\label{e3} \ee via function $Y (\rm x),$ which is obtained from
\emph{the Kerr theorem},
\cite{DKS,KraSte,Pen,PenRin,BurKerr,Multiks}. The PNC forms a
caustic at the Kerr singular ring, $r=\cos\theta=0 . $ As a
result, the aligned with Kerr PNC metric and electromagnetic
potential, \be A_{\m} = -P^{-2} Re \frac {e} { (r+ia \cos \theta)}
e^3_\m \label{Amu} , \ee
 concentrate near the
 Kerr ring, forming a gravitational waveguide -- a closed string
with the lightlike traveling waves \cite{Bur0,BurQ}.  Analysis of
the Kerr-Sen solution to low energy string theory \cite{KerSen}
showed that similarity of the Kerr ring with a closed strings is
not only analogue, but it has really the structure of a
fundamental heterotic string \cite{BurSen}.

Along with this closed string, the KN geometry contains also a
\emph{complex open string}, \cite{BurCStr}, which appears in the
initiated by Newman complex representation of Kerr geometry,
\cite{LinNew}.  This string gives an extra dimension $\theta$ to
the stringy source ($\theta \in [0, \pi] $), resulting in its
extension to a membrane (bubble source \cite{Lop,BurSol}. A
superstring counterpart of this extension is a transfer from
superstring theory to $11$-dimensional $M$-theory and $M2$-brane,
\cite{BBS}. In this essay we show that the adjoined to complex KN
string Kerr congruence is described as a quartic on the projective
twistor space $CP^3 ,$ realizing a Calabi-Yau twofold.
\medskip

\begin{figure}[ht]
\centerline{\epsfig{figure=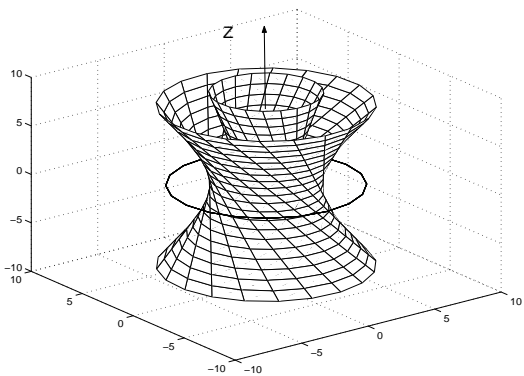,height=5cm,width=7cm}}
\caption{ Twistor null lines of the Kerr congruence are focused on
the Kerr singular ring, forming a twosheeted spacetime branched by
closed string.}
\end{figure}

 \textbf{Complex shift.} KN solution was initially
obtained by a "complex trick" \cite{KerNew}, and Newman
\cite{LinNew} showed
 that linearized KN solution may be generated by a
complex world line. This complex trick was first described by
Appel in 1887 \cite{App} as a \emph{complex shift}. Appel noticed
that Coulomb solution \be \phi(\vec x) = e/r = e/\sqrt{x^2+y^2
+z^2} \label{Coul}\ee is invariant under the shift $\vec x \to
\vec x +\vec a ,$ and considered complex shift of the origin,
$(x_0,y_0,z_0)=(0,0,0)$ along z-axis. $(x_0,y_0,z_0)= (0,0, -ia)
.$ On the real slice he obtained the complex potential
\begin{equation} \phi_a (\vec x) = Re \ e/{\tilde r} , \label{App}
\end{equation} with complex radial coordinate $\tilde r = r +ia \cos \theta .$
It was shown in \cite{Bur0,BurKerr,BurNst} that potential
(\ref{App}) corresponds exactly to KN electromagnetic field, and
the exact KN solution may be described as a field generated by a
\textbf{complex source propagating along complex world-line} \be
x_L^\m(\t_L) = x_0^\m (0) +  u^\m \t_L + \frac{ia}{2} \{ k^\m_L -
k^\m_R \} ,\label{cwL}\ee where $u^\m=(1,0,0,1), \quad
k_R=(1,0,0,-1), \quad k_L=(1,0,0,1).$ Index $L$ labels it as a
Left structure, and we should add a complex conjugate Right
structure \be x_R^\m(\t_R) = x_0^\m (0) +  u^\m \t_R -
\frac{ia}{2} \{ k^\m_L - k^\m_R \} .\label{cwR}\ee Therefore, from
complex point of view the Kerr and Schwarzschild geometries are
equivalent and differ only by their \emph{real slice}, which for
the Kerr solution goes aside of its center. Complex shift turns
the Schwarzschild radial directions $\vec n = \vec r /|r|$ into
twisted directions of the Kerr congruence, Fig.1.

\textbf{Kerr Theorem} determines shear free null congruence
(\ref{e3}) by the solution $Y(\rm x)$ of the equation \be F(T^A) =
0 \label{F0KerrTeor}, \ee where $F(T^A)$ is arbitrary holomorphic
function on the projective twistor space   \be T^A= \{ Y, \quad \l
^1 = \z - Y v, \quad \l ^2 =u + Y \Z \} .\label{(TA)} \ee

 Function $\bf F(T^A)$ for KN congruence is quadratic in $Y,$ and
\emph{\textbf{represents a quadric in $\bf CP^3 .$}} \noindent
Representing $F$ in coordinates $(Y,x^\m),$ we have
   $ F = A(x^\m)  Y^2 + B(x^\m) Y + C(x^\m), $ which yields two
   solutions \be Y^\pm (\rm x)= (- B \pm \D )/2A, \quad
 \tilde r = - \D = -(B^2 - 4AC)^{1/2} , \label{Ypm} \ee
allowing one to restore two PNC, (\ref{e3}).

\noindent \textbf{Complex string. } It was obtained
\cite{BurCStr,OogVaf} that the complex world line $x_0^\m (\t) ,$
parametrized by complex time $\t=t+i\sigma ,$ represents really a
two-dimensional surface which takes an intermediate position
between particle and string. The corresponding "hyperbolic string"
equation \cite{OogVaf},
 $\d_\t \d_{\bar\t}
x_0(t,\sigma) =0 ,$ yields the general solution \be x_0(t,\sigma)
= x_L(\t) + x_R(\bar\t) \ee as sum of the analytic and
anti-analytic modes $x_L(\t), \quad x_R(\bar\t),$ which are not
necessarily complex conjugate. For each real point $x^\m ,$ the
parameters $\t$ and $\bar\t$ should be determined by a complex
retarded-time construction. Complex source of the KN solution
corresponds to two \emph{straight} complex conjugate
world-lines,(\ref{cwL}),(\ref{cwR}). Contrary to the real case,
the complex retarded-advanced times $\t^\mp = t \mp \tilde r $ may
be determined by two different (Left or Right) complex null
planes, which are generators of the complex light cone. It yields
four different roots for the Left and Right complex structures
\cite{BurKerr,BurNst} \bea \t_L^\mp &=&
t \mp (r_L + ia\cos\theta_L) \label{Lretadv} \\
\t_R^\mp &=& t \pm (r_R + ia\cos\theta_R) \label{Rretadv}. \eea

The real slice condition determines relation $\sigma=a\cos \theta$
with  null directions  of the Kerr congruence $\theta \in [0, \pi]
,$ which puts restriction $\sigma \in [-a, a] $ indicating that
\emph{the complex string is open}, and  its endpoints $\sigma =
\pm a$ may be associated  with the Chan-Paton charges of a
quark-antiquark pair. In the real slice, the complex endpoints of
the string are mapped to the north and south twistor null lines,
$\theta =O,\pi ,$ see Fig.3.

\begin{figure}[ht]
\centerline{\epsfig{figure=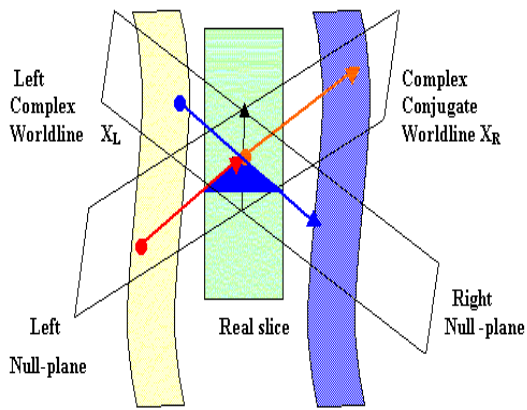,height=6cm,width=8.5cm}}
\caption{The complex conjugate Left and Right null planes generate
the Left and Right retarded and advanced roots.}
\end{figure}

\noindent\textbf{Orbifold.} The complex open string boundary
conditions \cite{BurCStr} require the \emph{worldsheet orbifold}
structure \cite{BBS,DHVW,HV,Hor,GSW} which
 turns the open string in a closed but
folded one. The world-sheet parity transformation $ \Omega: \sigma
\to - \sigma $ reverses orientation of the world sheet, and covers
it second time in mirror direction. Simultaneously, the Left and
Right modes are
 exchanged. \fn{Two oriented copies of the interval $\Sigma = [-a, a] ,$
$\Sigma^+ = [-a, a],$ and $ \Sigma^- = [-a, a]$ are joined,
forming a
 circle $ S^1 = \Sigma ^+\bigcup \Sigma ^-
,$ parametrized by $\theta ,$ and map $\theta \to \sigma=a\cos
\theta $ covers the world-sheet twice.} The projection $\Omega$ is
combined with space reflection $R: r\to - r ,$ resulting in
$R\Omega: \tilde r \to -\tilde r ,$ which relates the retarded and
advanced folds \be R\Omega:  \t^+ \to \t^- \label{POmt} ,\ee
preserving analyticity of the world-sheet.
\begin{figure}[ht]
\centerline{\epsfig{figure=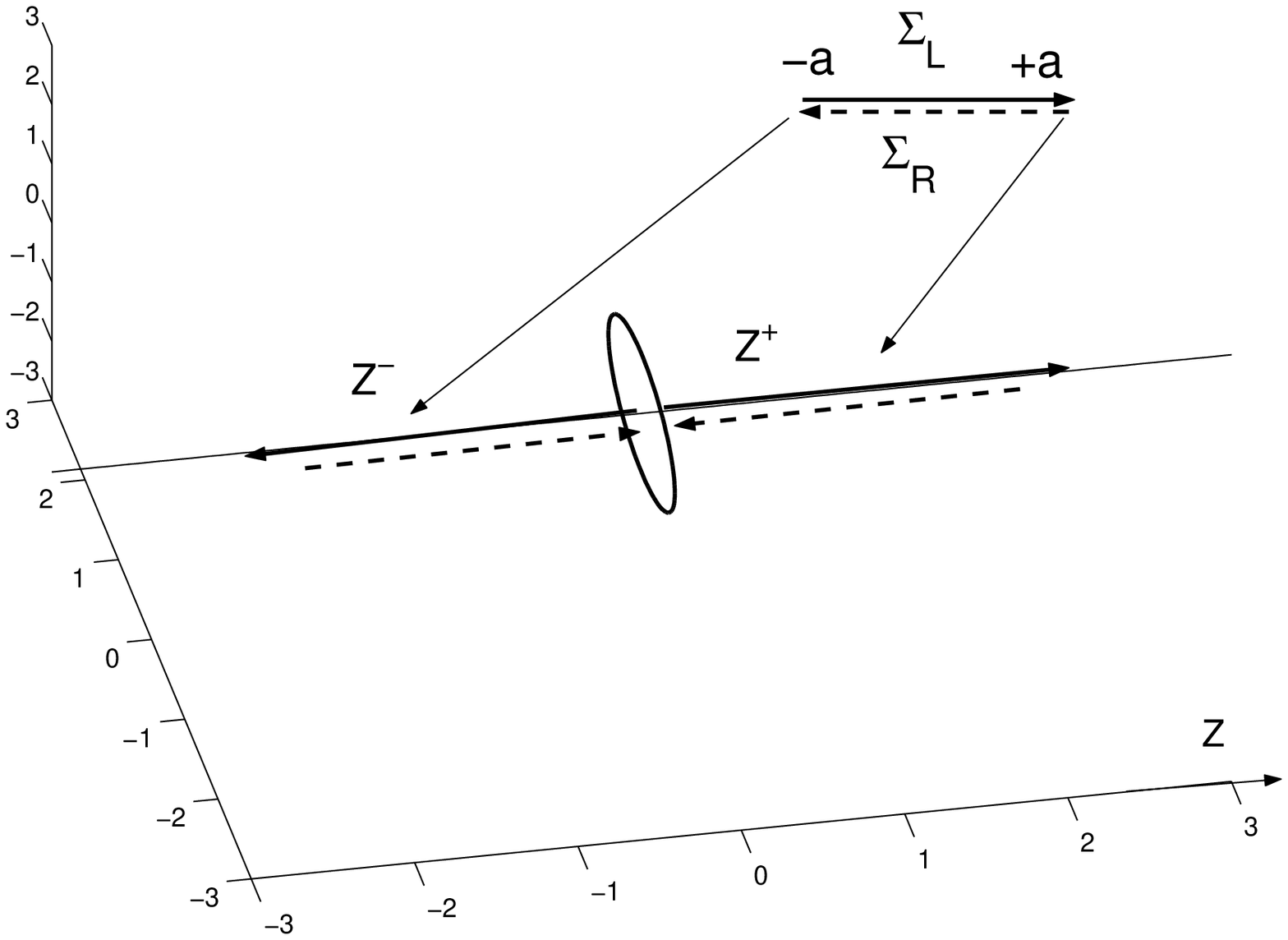,height=5cm,width=7cm}}
\caption{Ends of the open complex string, associated with quantum
numbers of quark-antiquark pair, are mapped onto the real
half-infinite $z^+, z^-$ axial strings. Dotted lines indicate
orbifold projection.}
\end{figure}
 The string modes $x_L(\t), \quad x_R(\bar\t),$ are extended on
the second half-cycle by the well known extrapolation,
\cite{BBS,GSW} \be x_L(\t^+) = x_R(\t^-); \quad x_R(\t^+) =
x_L(\t^-) , \label{orbi}\ee which forms the folded string, in with
the retarded and advanced modes are exchanged every half-cycle.

The real KN solution is generated by the straight complex
conjugate world lines (\ref{cwL}),(\ref{cwR}). Under string
excitations, the orbifold condition (\ref{orbi}) becomes
inconsistent with complex conjugation of the string ends, and
\emph{ the world lines $x_L(\t),$ $x_R(\bar\t)$ should be
considered as independent complex sources}. In accordance with
general treatment of the multiparticle Kerr-Schild solutions,
\cite{Multiks}, this corresponds to the Kerr theorem with a
two-particle generating function \be F_2(T^A) = F_L(T^A) F_R(T^A)
,\label{F2} \ee where $F_L$ and $F_R$ are determined by $x_L(\t),$
and $x_R(\bar\t),$ and both are quadratic in $T^A .$ The
corresponding equation \be F_2(T^A)=0 \ee describes \emph{a
quartic in $CP^3 $} which is the well known Calabi-Yau two-fold,
\cite{BBS,GSW}. We arrive at the result that excitations of the
Kerr complex string generate a Calabi-Yau two-fold on the
projective twistor space.

\textbf{Outlook.} One sees that the Kerr-Schild geometry displays
wonderful parallelism with superstring theory. In the recent paper
\cite{BurQ} we argued that it is not accidental, because
superstring theory contains gravity as its fundamental part, and
the KN gravity specifies the superstring structures. Along with
great similarity, the KN geometry  contains  very essential
differences:
\begin{itemize}
\item the Kerr strings live in four-dimensional space-time,
realizing a "compactification without compactification",
\cite{BurQ},

\item the Kerr strings are related with twistor theory and are
consistent with gravity by nature,

\item characteristic parameter of the Kerr strings $a=\hbar/m$
corresponds to Compton scale, which is closer to particle physics
vs. the Planck scale of superstring theory.

\end{itemize}
In many respects the Kerr-Schild gravity resembles the
twistor-string theory, \cite{Nair,Wit,BurTwi}, which is also
four-dimensional, based on twistors and related with experimental
physics. Taking also into account the holographic twistor
structure of the Kerr-Schild gravity, \cite{BurA,BurPreQ}, and its
close relationship with quantum world, one should treat it as a
specific prospective way to quantum gravity.

{\bf Acknowledgements.} Author thanks T. Nieuwenhuizen and
K.Stepaniants for useful discussions.

\end{document}